\documentclass{ws-ijmpe}
\usepackage{lscape}
\begin{document}
\title{Ground state properties and bubble structure of synthesized 
superheavy nuclei}
\author{\footnotesize S. K. Singh
\footnote{S. K. Singh, Email: shailesh@iopb.res.in}}
\address{Institute of Physics, Bhubaneswar-751 005, India.\\}

\author{\footnotesize M. Ikram
\footnote{M. Ikram, Email: ikram@iopb.res.in}}
\address{Department of Physics, Aligarh Muslim University, Aligarh-202002, India.\\}

\author{\footnotesize S. K. Patra
\footnote{Dr. S. K. Patra, Email: patra@iopb.res.in}}
\address{Institute of Physics, Bhubaneswar-751 005, India.\\}

\maketitle 







\begin{abstract}
We calculate the ground state properties of recently synthesized superheavy
elements from $Z = 105 - 118$ along with the predicted proton magic $Z = 120$.
The relativistic and nonrelativistic mean 
field formalisms are used to evaluate the binding energy, charge radius, 
quadrupole deformation parameter and the density distribution of nucleons. 
We analyzed the stability of the nuclei based on the binding energy and neutron 
to proton ratio. We also studied the bubble structure which
reveals the special features of the superheavy nuclei.
\end{abstract}

\section{Introduction}
The study of superheavy elements (SHEs) is an interesting topic for the current
days research in nuclear physics. Initially, the transuranium
elements were made by subjecting Uranium to a high neutron flux allowing
to capture neutrons successively. The neutron-rich isotopes undergo 
$\beta-$decay  and produce new elements. This process allows to prepare
element maximum upto Fermion ($Z=100$). This is due to the 
shorter time interval of the spontaneous fission than neutron capture time
of the newly form element. To 
make heavier $Z$ nuclei one has to reach the so called 
{\it island of stability} which is a matter of discussion from last five
decades \cite{goldhaber57,nilsson69,herrmann79,kumar89}. Earlier, 
it was predicted \cite{goldhaber57,myers66} that the next proton magic 
number beyond Z=82 would be 126 considering the equality of the proton 
and neutron magic numbers for known closed shell nuclei.  However, several 
microscopic calculations \cite{nilsson78,nilsson59,andersson76,meldner67,meldner78a,kohler71} 
suggest a shift of this number to 114. One of the cause of the shift is the 
Coulomb effect on the spherical single particle levels. The use of shell 
correction by V. M. Strutinsky \cite{strutinsky67} to the liquid-drop 
calculation of binding energy (BE) opens a more satisfactory
exploration towards the search of double close nucleus beyond $^{208}$Pb.
Using this approach, Z=114 is supported to be the proton magic after 82
in Refs.\cite{nix94,sob94,sob07,smo95}.
Based on the relativistic mean field (RMF) and Skyrme Hartree-Fock 
(SHF) calculations \cite{patra97,rutz97,patra99,meng07,bhuyan12}, it is 
predicted that the {\it island of stability} in the superheavy region 
is situated near mass number $A\sim 300$ peaking at $^{292,304}$120.

The main aim is to reach the {\it stability island} of superheavy valley
using some alternative process rather than the nucleosynthesis by neutron 
capture. Using cold fusion reaction, elements from $Z=107-112$ are synthesized
at GSI~\cite{hofman00,hofman81,hofman1995,hofman95,hofman96,hofman98,hofman09}
which are based on Pb and Bi targets. At the production time of Z = 112 
nucleus at GSI, the fusion cross section was extremely
small (1 pb) \cite{hofman96}, which led to the conclusion that
reaching still heavier elements will be very difficult by this process.
The element Z=113 was also synthesized in cold-fusion reaction
at RIKEN with a very low cross section $\sim 0.03$ pb \cite{morita07}
confirming the limitation of cold-fusion synthesis.

To overcome this problem in hot fusion evaporation reactions
with a deformed actinide targets and a neutron-rich doubly
magic spherical projectile like $^{48}Ca$ are used in the synthesis
of superheavy nuclei $Z = 112 - 118$ at 
Dubna~\cite{ogni98,ogni01,ogni04,ogni07,ogni10,ogni11}. 
It is worthwhile to mention that, although the synthesis of SHEs is 
a challenging task, the knowledge of their chemistry 
for the recently synthesized nuclei (Z=104-118) are more interesting to find 
the analogies between the properties of new elements and their lighter 
known neighbours in the groups of chemical elements and, eventually, proving 
their position in the periodic table. Nowadays, the chemistry of 
some trans-actinide elements, like Z=107, 108, 112 and 114 are known
experimentally \cite{turl02,dullmann02,eichler00,eichler07,eichler10} and theoretical efforts are
also given \cite{pers09}. These elements are located at the bottom of 
the periodic table. These nuclei are extremely unstable and the electron 
shells are influenced by strong relativistic effects 
\cite{frick69,frick71,pers02,pers05,barb11}.

The stability of a nucleus mostly depends on its structure which is again
a function of the combination of neutrons and protons number. 
Because of the large 
Coulombic repulsion, the central portion of the nucleus may take a 
depletion and create a bubble like structure. 
There are some interesting and extensive calculations
for various properties including bubbles or semi-bubbles structure of 
superheavy elements \cite{sobiczewski07,decharge99}.
However, theoretical predictions are in active discussion stage for the
ground state structure of these double close nuclei Z=114, 120 or 126 
and N=172 or 184. For example,
Z. Ren et al.\cite{ren01,ren02} predicted the nuclei have a well deformed 
ground state in RMF approach. Of course a spherical degenerate state is 
also available in this calculations. In contrast to the deform configuration 
of Ref. \cite{ren01,ren02}, Muntian et al. \cite{munt04} find a spherical shape 
in the macro-microscopic approach. In the frame-work of SHF with SkP or SLy7
interaction Cwiok et al \cite{cwio96} find $^{310}$126 as a spherical double
close shell nucleus.
In this paper, the stability and bubble structure of the superheavy nuclei 
are discussed by two well known self-consistent mean field models, namely
Skyrme Hartree-Fock (SHF) using SkI4 parameters \cite{reinhard95} with BCS 
pairing and Relativistic Mean Field (RMF) model with NL3* parameter set 
\cite{lalazissis09}.

The paper starts with a short introduction in Sec. I. The formalisms of 
SHF and RMF are presented in Sec.II. Mostly, the stability and structure 
of the synthesized nuclei will be discussed in Sec. III. The bubble 
properties will also be analyzed taking into account the density 
distribution of protons and neutrons. Finally a brief summary and
concluding remarks will be given in Sec. IV.

\section{Theoretical Formalisms}

\subsection{Skyrme Hartree-Fock (SHF) method}
There are many known parametrization of Skyrme interaction which reproduce the
experimental data for ground state properties of finite nuclei 
\cite{bender03,stone03} as well as the properties of infinite nuclear 
matter upto high density \cite{stone12}.
The general form of the Skyrme effective interaction
can be expressed as a density functional $\cal H$ with some 
empirical parameters \cite{bender03,chabanat1997,stone07}:
\begin{equation}
{\mathcal H}={\mathcal K}+{\mathcal H}_0+{\mathcal H}_3+ 
{\mathcal H}_{eff}+\cdots,
\label{eq:1}
\end{equation}
where ${\cal K}$ is the kinetic energy, ${\cal H}_0$ the zero range, ${\cal H}_3$ the
density dependent and ${\cal H}_{eff}$ the effective-mass dependent terms, which are
relevant for calculating the properties of nuclear matter.
More details can be found in Refs. \cite{bender03,chabanat1997,stone07}.
These are functions of 9 parameters
$t_i$, $x_i$ ($i=0,1,2,3$) and $\eta$ are given as
\begin{eqnarray}
{\mathcal H}_0&=&\frac{1}{4}t_0\left[(2+x_0)\rho^2 - (2x_0+1)(\rho_p^2+\rho_n^2)\right],
\label{eq:2}
\\
{\mathcal H}_3&=&\frac{1}{24}t_3\rho^\eta \left[(2+x_3)\rho^2 - (2x_3+1)(\rho_p^2+\rho_n^2)\right],
\label{eq:3}
\\
{\mathcal H}_{eff}&=&\frac{1}{8}\left[t_1(2+x_1)+t_2(2+x_2)\right]\tau \rho \nonumber \\
&&+\frac{1}{8}\left[t_2(2x_2+1)-t_1(2x_1+1)\right](\tau_p \rho_p+\tau_n \rho_n). \nonumber \\
\label{eq:4}
\end{eqnarray}
The kinetic energy ${\cal K}=\frac{\hbar^2}{2m}\tau$, a form used in the Fermi gas model for
non-interacting Fermions. The other terms ${\cal{H_{S\rho}}}$ and ${\cal{H_{S\vec J}}}$ with
$b_4$ and $b^{\prime}_4$ are the surface contributions which also responsible for a 
better spin-orbit interaction are defined as
\begin{eqnarray}
{\mathcal H}_{S\rho}&=&\frac{1}{16}\left[3t_1(1+\frac{1}{2}x_1)-t_2(1+\frac{1}{2}x_2)\right]
(\vec{\nabla}\rho)^2 \nonumber\\
&&-\frac{1}{16}\left[3t_1(x_1+\frac{1}{2})+t_2(x_2+\frac{1}{2})\right] \nonumber\\
&&\times\left[(\vec{\nabla}\rho_n)^2+(\vec{\nabla}\rho_p)^2\right],
\text{ and}
\label{eq:5}
\\
{\mathcal H}_{S\vec{J}}&=&-\frac{1}{2}\left[{b_4}\rho\vec{\nabla}\cdot\vec{J}+{b^{\prime}_4}(\rho_n\vec{\nabla}\cdot\vec{J_n}
+\rho_p\vec{\nabla}\cdot\vec{J_p})\right].
\label{eq:6}
\end{eqnarray}
Here, the total nucleon number density $\rho=\rho_n+\rho_p$, the kinetic energy density
$\tau=\tau_n+\tau_p$, and the spin-orbit density $\vec{J}=\vec{J}_n+\vec{J}_p$. The
subscripts $n$ and $p$ refer to neutron and proton, respectively, and $M$ is the nucleon
mass. The $\vec{J}_q=0$, $q=n$ or $p$, for spin-saturated nuclei, i.e., for nuclei
with major oscillator shells completely filled. The total binding energy (BE) of a
nucleus is the integral of the density functional $\cal H$.
In our calculations, we have used the Skyrme SkI4 parameter set \cite{reinhard95}.
This set is known to well account the spin-orbit interaction in finite nuclei,
related to the isotope shifts in Pb region and is better suited for the
study of exotic nuclei. Several more recent Skyrme parameters such as
SLy1-10, SkX, SkI5 and SkI6 are obtained by fitting the Hartree-Fock (HF) 
results with experimental data for nuclei starting from the valley of stability 
to neutron and proton drip-lines \cite{reinhard95,chabanat1997,chabanat98,brown98}.


\subsection{Relativistic mean field (RMF) formalism}

From last few decades, the RMF theory is applied successfully to study 
the structural properties of nuclei throughout the periodic 
table~\cite{serot,gam,ring,walecka,boguta97} starting from proton to neutron drip-lines.
The starting point of the RMF theory is the basic Lagrangian containing 
nucleons interacting with $\sigma-$, $\omega-$ and $\rho-$meson fields. The photon 
field $A_{\mu}$ is included to take care of the Coulomb interaction of protons. 
The relativistic mean field Lagrangian density is 
expressed as~\cite{serot,gam,ring,walecka,boguta97},

\begin{eqnarray}
{\cal L}&=&\bar{\psi_{i}}\{i\gamma^{\mu}
\partial_{\mu}-M\}\psi_{i}
+{\frac12}\partial^{\mu}\sigma\partial_{\mu}\sigma
-{\frac12}m_{\sigma}^{2}\sigma^{2}
-{\frac13}g_{2}\sigma^{3} \nonumber \\
&-&{\frac14}g_{3}\sigma^{4}-g_{s}\bar{\psi_{i}}\psi_{i}\sigma 
-{\frac14}\Omega^{\mu\nu}\Omega_{\mu\nu}+{\frac12}m_{w}^{2}V^{\mu}V_{\mu}\nonumber \\
&-&g_{w}\bar\psi_{i}\gamma^{\mu}\psi_{i}
V_{\mu}-{\frac14}\vec{B}^{\mu\nu}\vec{B}_{\mu\nu} 
+{\frac12}m_{\rho}^{2}{\vec{R}^{\mu}}{\vec{R}_{\mu}}-{\frac14}F^{\mu\nu}F_{\mu\nu} \nonumber\\
&-&g_{\rho}\bar\psi_{i}\gamma^{\mu}\vec{\tau}\psi_{i}\vec{R^{\mu}}-e\bar\psi_{i}
\gamma^{\mu}\frac{\left(1-\tau_{3i}\right)}{2}\psi_{i}A_{\mu} .
\end{eqnarray}
Here M, $m_{\sigma}$, $m_{\omega}$ and $m_{\rho}$ are the masses for nucleon, 
${\sigma}$-, ${\omega}$- and ${\rho}$-mesons and ${\psi}$ is its Dirac spinor. 
The field for the ${\sigma}$-meson is denoted by ${\sigma}$, ${\omega}$-meson 
by $V_{\mu}$ and ${\rho}$-meson by $R_{\mu}$. 
$g_s$, $g_{\omega}$, $g_{\rho}$ and $e^2/4{\pi}$=1/137 are the coupling 
constants for the ${\sigma}$, ${\omega}$, ${\rho}$-mesons and photon respectively.
$g_2$ and $g_3$ are the self-interaction coupling constants for ${\sigma}$ mesons.
By using the classical variational principle we obtain the field equations for 
the nucleons and mesons. 
\begin{eqnarray}
\{-\bigtriangleup+m^2_\sigma\}\sigma^0(r_{\perp},z)&=&-g_\sigma\rho_s(r_{\perp},z)\nonumber\\
&-& g_2\sigma^2(r_{\perp},z)-g_3\sigma^3(r_{\perp},z) ,\\
\{-\bigtriangleup+m^2_\omega\}V^0(r_{\perp},z)&=&g_{\omega}\rho_v(r_{\perp},z) ,\\
\{-\bigtriangleup+m^2_\rho\}R^0(r_{\perp},z)&=&g_{\rho}\rho_3(r_{\perp},z) , \\
-\bigtriangleup A^0(r_{\perp},z)&=&e\rho_c(r_{\perp},z). 
\end{eqnarray}
The Dirac equation for the nucleons is written by
\begin{equation}
\{-i\alpha.\triangledown + V(r_{\perp},z)+\beta M^\dagger\}\psi_i=\epsilon_i\psi_i.
\end{equation}
The effective mass of the nucleon is
\begin{equation}
M^\dagger=M+S(r_{\perp},z)=M+g_\sigma\sigma(r_{\perp},z),
\end{equation}
and the vector potential is
\begin{equation}
 V(r_{\perp},z)=g_{\omega}V^{0}(r_{\perp},z)+g_{\rho}\tau_{3}R^{0}(r_{\perp},z)+
e\frac{(1-\tau_3)}{2}A^0(r_{\perp},z). 
\end{equation}
A static solution is obtained from the equations of motion to describe 
the ground state properties of nuclei.
The set of nonlinear coupled equations are solved self-consistently in an axially
deformed harmonic oscillator basis $N_F=N_B=20$. 
The quadrupole deformation parameter $\beta_{2}$ is extracted from the 
calculated quadrupole moments of neutrons and protons through 
\begin{equation}
Q = Q_n + Q_p = \sqrt{\frac{16\pi}5} \left(\frac3{4\pi} AR^2\beta_2\right),
\end{equation}
where R = 1.2$A^{1/3}$.\\
The total energy of the system is given by 
\begin{equation}
 E_{total} = E_{part}+E_{\sigma}+E_{\omega}+E_{\rho}+E_{c}+E_{pair}+E_{c.m.},
\end{equation}
where $E_{part}$ is the sum of the single particle energies of the nucleons and 
$E_{\sigma}$, $E_{\omega}$, $E_{\rho}$, $E_{c}$, $E_{pair}$, $E_{cm}$ are 
the contributions of the meson fields, the Coulomb field, pairing energy  
and the center-of-mass energy, respectively.
We use the recently reported NL3* parameter set~\cite{lalazissis09} in our 
calculations for RMF formalism. 


\section{Results and discussions}

In the present paper, we calculate the binding energy, root mean square (rms)
charge radius $r_{ch}$ and quadrupole deformation parameter $\beta_2$. A special 
attention is given to the neutron and proton density distributions of these
superheavy nuclei, where we search for the bubble structure at various 
excited intrinsic states. The results are explained in subsection 3.1 to 3.5.

\subsection{Stability of nuclei}

While analyzing the nuclear landscape, the stability of nucleus is maximum
with a neutron to proton ratio $N/Z$ at about 1 for light mass region. This
value of $N/Z$ increases with mass number A to neutralize the Coulomb repulsion
and saturated at $\sim 1.5$ for naturally occuring nuclei.
For example, the neutron to proton ratio for $^{208}$Pb is 1.537 and
for $^{235,238}$U, these values are 1.554 and 1.587, respectively. 
To see the
overall N/Z values of the synthesized isotopes in the superheavy region, we 
calculate the neutron to proton ratio for $Z$=105 to 118 in addition to 
the predicted double close nuclei $^{292,293,304}$120. 
We find the minimum and maximum $N:Z$ as 1.433 and 1.614, respectively.

\begin{figure}[h]
\vspace{0.75cm}
\includegraphics[width=0.95\columnwidth]{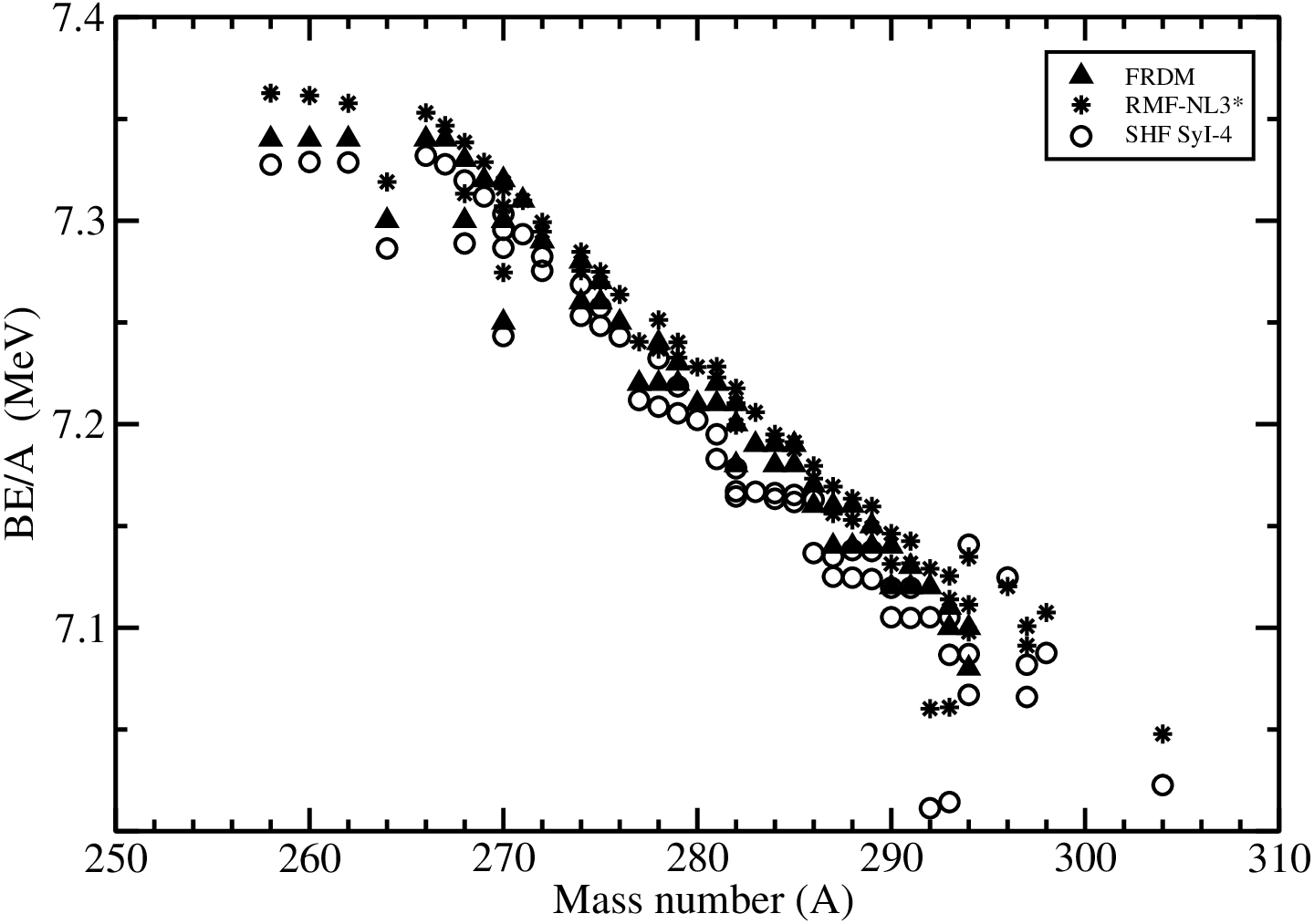}
\caption{\label{be}The binding energy per nucleon $(BE/A)$ is compared with the
FRDM data. The triangles, stars and circles are for FRDM, RMF(NL3*) and 
SHF(SkI4), respectively. }
\label{fig1}
\end{figure}

One important quantity is the binding energy $(BE)$ of a nucleus. 
The stability of a nucleus does not fully depend on the binding energy only. 
The arrangement of shells inside the nucleus is also a function of stability. 
In the absence of shell energy contribution (liqiud-drop limit), it may be possible that 
a general view on stability of nuclei can be made on the basis 
of binding energy per particle $(BE/A)$, i.e., nucleus is more 
stable with greater value of  $BE/A$. Binding energy is one of the experimental 
observable, which can be measured precisely and could be an important input 
for stability. Generally, a nucleus is more stable with a larger binding energy 
per particle along an isotopic chain. The trend of $BE/A$ as a 
function of mass number A is shown in Figure ~\ref{be}. It is clear that low mass 
nuclei in the vicinity of $\beta-$stability valley have more life-time compare 
to the heavier isotopes. 
This may be due to the large Coulomb repulsion of the superheavy nuclei,
although the $N/Z$ ratio is almost constant. The declining trend of $BE/A$ 
indicates the difficulty in the formation of heavier superheavy nuclei. 
As we have mentioned, not only the BE, but also the internal shell structure
plays for the role of stability. Thus, there is a bright chance for the formation
of superheavy nucleus if a proper combination of N and Z are selected which
could be the next double close nucleus. In this context, it is to be recalled that 
the formation of a neutron-proton bound 
state system requires only 2.224 MeV which is the typical binding energy 
of Deuteron ($BE/A\sim 1.112$ MeV) \cite{audi03}.  


\subsection{Binding energy and quadrupole deformation parameter}

The ground state binding energy and quadrupole deformation 
parameter for all the observed nuclei considered here are 
given in Tables I and II. The much discussed superheavy isotopes 
$^{292,304}$120 which are considered
to be the next double closed nuclei beyond $^{208}$Pb are also listed.
To find the ground state solution, the calculations are done
with an initial spherical, prolate and oblate quadrupole deformation parameter
$\beta_0$ both in the Skyrme Hartree-Fock (SHF) and relativistic
mean field (RMF) formalisms. The well known SkI4\cite{reinhard95} 
and NL3*\cite{lalazissis09} parametrizations are used through out the
SHF and RMF calculations, respectively. It is to be noted that the
maximum binding energy corresponds to the ground state and all other solutions
are the intrinsic excited state configurations. Considering this criteria, 
we evaluate
the ground state binding energy and also the first and second excited
states wherever available. The results obtained from SHF are listed in Table I
and those for RMF are listed in Table II. Since the experimental binding 
energy of these nuclei are not known, to get a first hand information
about the predictive power of our calculations, we have inserted the most
commonly used binding energy of finite range droplet model (FRDM) 
\cite{moller97}. The quadrupole deformation parameter $\beta_2$
obtained in FRDM is also given for comparison \cite{moller95}. 
Comparing Tables I and II, it is remarkable to note that the binding
energy obtained by SHF(SkI4) and RMF(NL3*) are quite agreeable 
with FRDM prediction.  The deviation of our
results (from the FRDM prediction) is a maximum of about $0.5\%$. A further 
inspection of the results finds an edge of RMF(NL3*) over SHF(SkI4).
In general, the calculated binding energies and deformation parameters are 
in similar trend with each other in all the three models.


\begin{landscape}
\begin{table}
\caption{The BE (MeV), $\beta_2$, $r_{ch}$ (fm) (ground state) and D.F. (\%) using SHF(SkI4).}
\renewcommand{\tabcolsep}{0.08cm}
\renewcommand{\arraystretch}{0.5}
\begin{tabular}{|c|c|c|c|c|c|c|c|c|c|c|c|c|c|c|c|}
\hline\hline
Nuclei&\multicolumn{3}{c|}{BE}&\multicolumn{3}{c|}{$\beta_2$}&$r_{ch.}$&\multicolumn{2}{c|}{sph.}
&\multicolumn{2}{c|}{prol.}&\multicolumn{2}{c|}{obl.}&\multicolumn{2}{c|}{FRDM}\\
\hline
&sph.&prol.&obl.&sph.&prol.&obl.&&(DF)$_n$&(DF)$_p$&(DF)$_n$&(DF)$_p$&(DF)$_n$&(DF)$_p$&BE&$\beta_2$\\
\hline
$^{266}$Db&& 1950.3& 1939.7&& 0.262& -0.282& 6.08 &0.0   &0.0  &0.0 &2.98&1.32  &7.86  &1953.42 &0.221\\
$^{267}$Db&& 1956.5& 1945.9&& 0.258& -0.283& 6.09 &0.0   &3.41 &0.0 &3.41&1.29  &7.43  &1960.49 &0.230\\
$^{268}$Db&& 1961.7& 1951.3&& 0.254& -0.193& 6.09 &0.0   &3.42 &0.0 &3.83&1.35  &7.36  &1965.86 &0.221\\
$^{269}$Db&& 1966.9& 1957.6&& 0.250& -0.191& 6.10 &0.0   &3.58 &0.0 &3.60&1.13  &7.53  &1972.01 &0.221\\
$^{270}$Db&& 1971.9& 1963.7&& 0.259& -0.188& 6.11 &0.0   &3.65 &0.0 &2.12&0.81  &7.64  &1976.81 &0.212\\
$^{258}$Sg&& 1890.5& 1880.0&& 0.287& -0.287& 6.05 &2.73  &4.52 &2.73&4.54&2.74  &0.89  &1895.22 &0.248\\
$^{260}$Sg&& 1905.5& 1894.9&& 0.284& -0.286& 6.06 &2.29  &4.24 &0.0 &4.24&1.75  &12.86 &1909.90 &0.239\\
$^{262}$Sg&& 1920.1& 1909.3&& 0.279& -0.283& 6.07 &1.05  &3.69 &1.72&3.70&1.59  &12.19 &1923.94 &0.229\\
$^{271}$Sg&& 1976.5& 1968.0&& 0.254& -0.188& 6.12 &1.39  &4.12 &0.0 &3.45&0.93  &7.71  &1981.87 &0.212\\
$^{270}$Bh&& 1969.8& 1958.7&& 0.254& -0.192& 6.11 &0.80  &4.00 &0.0 &4.02&1.53  &7.57  &1973.53 &0.222\\
$^{272}$Bh&& 1980.8& 1971.9&& 0.244& -0.187& 6.12 &0.0   &4.21 &0.0 &4.62&1.07  &7.82  &1985.27 &0.221\\
$^{274}$Bh&& 1987.4& 1984.7&& 0.237& -0.182& 6.13 &0.0   &4.25 &0.0 &3.16&0.52  &8.20  &1996.40 &0.192\\
$^{264}$Hs&& 1923.6& 1912.8&& 0.271& -0.283& 6.08 &0.0   &4.42 &1.23&3.74&1.36  &12.27 &1927.62 &0.229\\
$^{268}$Hs&& 1953.4& 1941.3&& 0.262& -0.281& 6.12 &0.0   &3.60 &0.0 &4.40&1.26  &11.46 &1957.28 &0.230\\
$^{270}$Hs&& 1967.4& 1955.2&& 0.256& -0.193& 6.11 &0.0   &4.17 &0.0 &4.17&1.76  &7.69  &1971.42 &0.231\\
$^{272}$Hs&& 1978.9& 1968.8&& 0.248& -0.188& 6.12 &0.0   &4.40 &0.0 &4.40&1.48  &7.84  &1984.20 &0.222\\
$^{275}$Hs&& 1995.8& 1988.7&& 0.233& -0.180& 6.14 &0.0   &4.39 &0.0 &3.84&0.78  &8.42  &2000.78 &0.183\\
$^{274}$Mt&& 1987.4& 1978.7&& 0.237& -0.183& 6.14 &0.0   &4.71 &0.0 &4.72&1.49  &8.28  &1991.77 &0.222\\
$^{275}$Mt&& 1993.3& 1985.5&& 0.232& -0.181& 6.14 &0.0   &4.31 &0.0 &4.73&1.27  &8.49  &1998.36 &0.212\\
$^{276}$Mt&& 1999.1& 1992.2&& 0.225& -0.178& 6.14 &0.0   &4.27 &0.0 &4.67&1.05  &8.73  &2003.64 &0.202\\
$^{278}$Mt&& 2010.6& 2005.5&& 0.212& -0.173& 6.15 &0.0   &4.38 &0.0 &4.49&0.70  &9.30  &2015.32 &0.136\\
$^{270}$Ds&& 1955.7& 1945.5&& 0.250& -0.195& 6.12 &0.0   &4.30 &0.0 &4.31&2.09  &8.13  &1960.53 &0.221\\
$^{279}$Ds& 2014.0& 2011.7& 2009.1& 0.199& 0.276& -0.171& 6.20 &0.0   &4.29 &0.0 &4.81&1.08  &9.70  &2019.39 &0.127\\
$^{281}$Ds& 2019.8& 2021.8& 2022.3& -0.030& 0.340& -0.166& 6.24&16.96 &22.10&0.0 &4.44&0.88  &10.34 &2031.91 &0.108\\
$^{277}$Rg&& 1997.7& 1991.3&& 0.220& -0.176& 6.16 &0.0   &4.78 &0.0 &4.32&2.03  &9.38  &2003.04 &0.193\\
$^{278}$Rg&& 2004.0& 1998.4&& 0.220& -0.173& 6.17 &0.0   &4.12 &0.0 &4.72&1.86  &9.63  &2008.68 &0.164\\
$^{279}$Rg&& 2007.1& 2005.4&& 0.288& -0.170& 6.21 &0.0   &4.09 &0.8 &4.24&1.68  &9.90  &2015.73 &0.164\\
$^{280}$Rg&& 2012.5& 2012.3&& 0.334& -0.167& 6.24 &0.0   &4.11 &0.0 &4.83&1.53  &10.19 &2021.81 &0.117\\
$^{281}$Rg& 2016.9& 2018.4& 2019.2& -0.036& 0.353& -0.165& 6.26&17.27 &22.44&0.77&4.56&1.40  &10.50 &2028.85 &0.099\\
$^{282}$Rg& 2023.8& 2024.3& 2025.9& -0.037& 0.366& -0.163& 6.27&17.06 &20.14&0.62&4.61&1.29  &10.79 &2034.82 &0.099\\
$^{282}$112& 2020.4& 2020.4& 2022.3& -0.030& 0.370& -0.161& 6.28&17.85&23.20&0.0 &4.75&1.94  &11.07 &2032.68 &0.089\\
$^{283}$112& 2027.9& 2028.2& 2029.2& -0.040& 0.489& -0.159& 6.38&16.73&22.14&0.0 &3.99&1.29  &11.33 &2038.67 &0.089\\
$^{284}$112& 2035.1& 2035.2& 2036.0& -0.066& 0.525& -0.159& 6.41&14.18&20.42&1.41&2.78&2.34  &11.93 &2045.88 &0.089\\
$^{285}$112& 2042.1& 2041.6& 2042.5& -0.068& 0.532& -0.167& 6.43&13.29&19.93&0.0 &2.47&1.16  &11.20 &2051.61 &0.089\\
$^{294}$112& 2099.4& 2093.3& 2098.9& -0.004& 0.559& -0.005& 6.50&3.07 &15.17&0.0 &5.74&0.22  &11.86 &2103.92 &0.00 \\

\hline
continue....
\end{tabular}
\end{table}
\end{landscape}

\begin{landscape}
\begin{table}
\renewcommand{\tabcolsep}{0.08cm}
\renewcommand{\arraystretch}{0.5}
\begin{tabular}{|c|c|c|c|c|c|c|c|c|c|c|c|c|c|c|c|}
\hline
Nuclei&\multicolumn{3}{c|}{BE}&\multicolumn{3}{c|}{$\beta_2$}&$r_{ch.}$&\multicolumn{2}{c|}{sph.}
&\multicolumn{2}{c|}{prol.}&\multicolumn{2}{c|}{obl.}&\multicolumn{2}{c|}{FRDM}\\
\hline
&sph.&prol.&obl.&sph.&prol.&obl.&&(DF)$_n$&(DF)$_p$&(DF)$_n$&(DF)$_p$&(DF)$_n$&(DF)$_p$&BE&$\beta_2$\\
\hline
$^{282}$113& 2021.0& 2021.1& 2017.9& -0.164& 0.171& -0.159& 6.18&1.02 &9.57 &0.0 &8.43&2.71  &11.44 &2027.20 &0.071\\
$^{284}$113& 2031.0& 2034.4& 2032.3& -0.002& 0.158& -0.154& 6.18&18.47&24.09&0.0 &8.61&2.37  &11.93 &2040.95 &0.080\\
$^{285}$113& 2038.2& 2041.1& 2039.2& -0.023& 0.151& -0.155& 6.18&18.02&23.72&0.0 &8.79&2.06  &11.90 &2048.18 &0.072\\
$^{286}$113& 2045.3& 2047.7& 2045.9& -0.021& 0.144& -0.163& 6.19&17.54&23.44&0.0 &8.94&1.58  &11.65 &2054.22 &0.072\\
$^{286}$114& 2041.4& 2041.1& 2042.0& -0.012& 0.555& -0.150& 6.46&18.59&24.53&0.0 &0.0 &2.63  &12.55 &2051.59 &-0.096\\
$^{287}$114& 2048.6& 2047.7& 2048.8& -0.012& 0.554& -0.158& 6.46&18.06&24.19&0.0 &4.82&2.06  &12.13 &2057.65 &-0.078\\
$^{288}$114& 2055.8& 2054.1& 2055.5& -0.024& 0.547& -0.165& 6.46&16.76&23.18&0.0 &4.63&2.14  &12.09 &2065.01 &0.053 \\
$^{289}$114& 2062.8& 2060.3& 2062.2& -0.011& 0.537& -0.149& 6.46&16.07&22.95&0.0 &4.81&2.23  &12.36 &2071.04 &-0.052\\
$^{296}$114& 2108.9& 2102.3& 2108.4& -0.003& 0.565& -0.043& 6.53&4.05 &16.58&0.0 &4.32&1.68  &13.68 &2113.22 &0.00  \\
$^{298}$114& 2121.1& 2113.3& 2120.8&  0.000& 0.566& -0.012& 6.54&0.0  &11.18&0.0 &4.72&0.0   &10.74 &2123.30 &0.00  \\
$^{287}$115& 2043.1& 2044.9& 2044.1& -0.008& 0.153& -0.137& 6.20&19.86&25.48&0.0 &6.57&4.18  &13.69 &2052.72 &-0.096\\
$^{288}$115& 2050.5& 2051.9& 2051.2& -0.005& 0.145& -0.137& 6.20&19.24&25.09&0.0 &7.07&3.66  &13.38 &2059.12 &-0.087\\
$^{289}$115& 2057.8& 2058.8& 2057.9& -0.002& 0.137& -0.141& 6.21&18.19&25.44&0.0 &7.48&3.02  &12.92 &2066.45 &0.053 \\
$^{290}$115& 2064.9& 2064.7& 2065.2&  0.000& 0.117& -0.119& 6.20&16.88&23.61&2.34&10.75&3.83 &13.59 &2072.59 &-0.079\\
$^{291}$115& 2071.9& 2071.8& 2072.4&  0.000& 0.064& -0.102& 6.20&15.46&22.70&8.62&19.14&4.39 &14.44 &2079.83 &-0.061\\
$^{290}$116& 2059.5& 2060.5& 2060.3&  0.002& 0.134& -0.121& 6.21&17.92&24.59&0.0 &7.90 &4.92 &15.51 &2068.76 &0.072 \\
$^{291}$116& 2066.7& 2066.7& 2067.5&  0.002& 0.110& -0.108& 6.21&16.61&23.78&0.3 &11.81&5.56 &15.29 &2074.84 &0.072 \\
$^{292}$116& 2073.9& 2074.1& 2074.7&  0.002& 0.062& -0.097& 6.21&15.32&22.96&10.18&19.89&5.37&15.45 &2082.48 &-0.070\\
$^{293}$116& 2080.9& 2081.1& 2081.8&  0.002& 0.052& -0.092& 6.21&13.55&21.98&10.76&2.35&4.07 &14.79 &2088.40 &-0.070\\
$^{293}$117& 2075.6& 2074.3& 2076.4&  0.039& 0.536& -0.099& 6.50&13.64&22.17&0.0 &0.0 &4.57  &14.65 &2083.06 &-0.087\\
$^{294}$117& 2082.8& 2080.9& 2083.6&  0.043& 0.538& -0.095& 6.51&12.18&21.27&0.0 &0.0 &2.99  &13.77 &2089.22 &-0.087\\
$^{297}$117& 2103.6& 2100.6& 2103.3&  0.031& 0.561& -0.083& 6.55&8.27 &18.36&0.0 &0.0 &0.0   &11.86 &2109.27 &-0.008\\
$^{294}$118& 2076.9& 2076.1& 2077.7&  0.055& 0.548& -0.102& 6.52&9.32 &19.71&0.0 &0.34&3.74  &13.73 &2084.78 &-0.087\\
$^{297}$118& 2098.2& 2096.6& 2098.6& -0.023& 0.561& -0.087& 6.55&11.38&20.22&0.0 &1.28&0.0   &11.30 &2104.59 &-0.035\\
$^{292}$120& 2046.4& 2047.1& 2047.3& -0.006& 0.530& -0.125& 6.50&26.49&31.54&0.4 &0.0 &7.58  &15.13 &2055.19 &-0.130\\
$^{293}$120& 2054.5& 2054.9& 2055.2& -0.005& 0.528& -0.124& 6.51&25.35&30.75&0.3 &0.98&6.71  &14.56 &2062.42 &0.089 \\
$^{304}$120& 2134.9& 2132.5& 2132.7& -0.008& 0.558& -0.074& 6.59&0.0  &8.36 &0.0 &3.62&0.0   &10.01 &2137.99 &0.00  \\
\hline
\end{tabular}
\end{table}
\end{landscape}

\begin{table*}
\caption{Same as Table I with  RMF(NL3*) model.}
\renewcommand{\tabcolsep}{0.08cm}
\renewcommand{\arraystretch}{0.5}
\begin{tabular}{|c|c|c|c|c|c|c|c|c|c|c|c|c|c|}
\hline\hline
Nuclei&\multicolumn{3}{c|}{BE}&\multicolumn{3}{c|}{$\beta_2$}&$r_{ch.}$&\multicolumn{2}{c|}{sph.}
&\multicolumn{2}{c|}{prol.}&\multicolumn{2}{c|}{obl.}\\
\hline
&sph.&prol.&obl.&sph.&prol.&obl.&&(DF)$_n$&(DF)$_p$&(DF)$_n$&(DF)$_p$&(DF)$_n$&(DF)$_p$\\
\hline
$^{266}$Db& 1941.5& 1955.9& 1946.2& 0.002& 0.273& -0.308&6.129& 19.57& 18.00& 1.22& 2.95& 5.22& 20.05\\
$^{267}$Db& 1947.9& 1961.6& 1952.1& 0.002& 0.269& -0.307&6.133& 19.79& 18.11& 1.04& 1.91& 5.28& 15.67\\
$^{268}$Db& 1954.3& 1966.7& 1957.9& 0.001& 0.267& -0.306&6.138& 19.98& 18.29& 0.92& 2.92& 4.08& 20.32\\
$^{269}$Db& 1960.4& 1971.4& 1963.6& 0.001& 0.263& -0.306&6.143& 20.20& 18.52& 0.87& 1.55& 4.19& 20.66\\
$^{270}$Db& 1966.5& 1976.0& 1969.1& 0.002& 0.256& -0.306&6.147& 20.40& 18.78& 0.87& 1.23& 4.27& 21.14\\
$^{258}$Sg& 1883.6& 1899.6& 1889.5& 0.004& 0.291& -0.301&6.091& 17.28& 16.26& 6.26& 4.43& 9.54& 18.95\\
$^{260}$Sg& 1898.1& 1913.9& 1903.9& 0.002& 0.285& -0.301&6.102& 18.04& 16.86& 6.18& 4.46& 7.35& 19.66\\
$^{262}$Sg& 1912.1& 1927.7& 1917.7& 0.002& 0.281& -0.305&6.113& 18.68& 17.38& 5.32& 4.33& 5.69& 19.37\\
$^{271}$Sg& 1971.5& 1980.9& 1973.6& 0.002& 0.254& -0.309&6.157& 20.68& 19.31& 0.38& 3.48& 5.15& 20.99\\
$^{270}$Bh& 1963.1& 1975.3& 1965.9& 0.002& 0.262& -0.311&6.158& 0.42 & 19.16& 0.60& 4.21& 4.58& 20.16\\
$^{272}$Bh& 1975.9& 1985.4& 1977.8& 0.002& 0.244& -0.312&6.164& 20.87& 19.69& 0.41& 4.11& 4.89& 20.86\\
$^{274}$Bh& 1995.9& 1990.8& 1989.3& 0.205& 0.216& -0.319&6.162& 1.00 & 6.77& 0.54& 0.26& 5.99& 22.15\\
$^{264}$Hs&& 1932.1& 1922.6&& 0.269& -0.311&6.130& 5.72 & 7.62& 6.06& 4.62& 5.28& 19.75\\
$^{268}$Hs&& 1959.9& 1950.1&& 0.265& -0.315&6.152& 1.98 & 5.76& 1.86& 3.08& 3.64& 19.94\\
$^{270}$Hs&& 1972.9& 1963.1&& 0.260& -0.314&6.161& 0.78 & 5.24& 0.88& 4.85& 4.34& 20.10\\
$^{272}$Hs&& 1984.1& 1975.6&& 0.249& -0.313&6.169& 0.43 & 5.29& 0.99& 4.78& 3.55& 20.51\\
$^{275}$Hs&& 2000.6& 1993.6&& 0.203& -0.322&6.171& 1.19 & 7.66& 1.40& 8.53& 5.89& 22.19\\
$^{274}$Mt&& 1993.4& 1985.1&& 0.216& -0.319&6.173& 1.08 & 7.69& 1.49& 1.16& 5.09& 21.46\\
$^{275}$Mt&& 1999.1& 1991.3&& 0.207& -0.324&6.176& 1.22 & 8.00& 1.47& 0.96& 5.69& 22.16\\
$^{276}$Mt&& 2004.7& 1997.4&& 0.199& -0.327&6.180& 1.47 & 8.16& 1.43& 8.98& 6.16& 22.77\\
$^{278}$Mt& 2015.8& 2013.6& 2009.4& 0.185& 0.373& -0.330&6.186& 1.39 & 8.30& 1.06& 3.68& 6.53& 23.77\\
$^{270}$Ds&& 1963.9& 1954.9&& 0.251& -0.319&6.167& 2.84 & 7.77& 2.88& 8.09& 4.35& 20.93\\
$^{279}$Ds& 2020.0& 2017.4& 2013.3& 0.182& 0.384& -0.336&6.196& 1.40 & 8.57& 2.63& 3.88& 6.81& 24.27\\
$^{281}$Ds& 2031.2& 2029.2& 2025.4& 0.168& 0.476& -0.334&6.202& 1.65 & 8.86& 4.93& 7.06& 6.76& 24.47\\
$^{277}$Rg& 2005.6& 2002.7& 1997.8& 0.199& 0.341& -0.345&6.194& 1.39 & 8.61& 2.16& 6.57& 7.15& 25.13\\
$^{278}$Rg& 2011.8& 2008.8& 2004.2& 0.191& 0.349& -0.345&6.198& 1.41 & 8.69& 2.23& 6.55& 7.18& 25.22\\
$^{279}$Rg& 2017.9& 2014.7& 2010.7& 0.185& 0.362& -0.345&6.202& 1.41 & 8.73& 2.38& 6.58& 7.18& 25.26\\
$^{280}$Rg& 2023.9& 2020.6& 2017.0& 0.179& 0.385& -0.343&6.206& 1.41 & 8.79& 2.64& 6.42& 7.16& 25.27\\
$^{281}$Rg& 2029.7& 2027.2& 2023.3& 0.173& 0.503& -0.341&6.209& 1.46 & 8.89& 6.42& 8.57& 7.09& 25.22\\
$^{282}$Rg& 2035.4& 2033.3& 2029.4& 0.166& 0.497& -0.339&6.212& 1.59 & 9.04& 5.86& 8.17& 7.05& 25.23\\
$^{282}$112& 2033.3& 2030.8& 2026.8& 0.171& 0.515& -0.346&6.219& 1.45  & 9.06 & 6.79& 8.79 & 7.33& 25.95\\
$^{283}$112& 2039.2& 2037.2& 2033.2& 0.165& 0.508& -0.343&6.222& 1.53  & 9.17 & 6.22& 7.65 & 7.29& 25.92\\
$^{284}$112& 2041.3& 2043.4& 2039.3& 0.002& 0.501& -0.340&6.436& 23.96 & 22.74& 5.72& 8.33 & 7.25& 25.95\\
$^{285}$112& 2047.4& 2049.5& 2045.3& 0.002& 0.496& -0.337&6.438& 23.56 & 22.59& 5.38& 8.38 & 7.15& 26.04\\
$^{294}$112& 2097.6& 2097.3& 2098.6& 0.001& 0.516& -0.024&6.240& 8.94  & 17.06& 6.11& 9.59 & 7.89& 17.18\\
$^{282}$113& 2030.3& 2025.4& 2023.6& 0.175& 0.363& -0.352&6.226& 1.42  & 9.04 & 2.27& 7.30 & 7.45& 26.41\\
$^{284}$113& 2041.9& 2042.5& 2036.5& 0.139& 0.164& -0.347&6.233& 4.71  & 11.34& 1.55& 1.03 & 7.46& 26.42\\
$^{285}$113& 2045.1& 2048.5& 2042.8& 0.001& 0.158& -0.344&6.236& 24.18 & 22.89& 2.05& 10.44& 7.46& 26.48\\
$^{286}$113& 2051.4& 2053.4& 2051.5& 0.002& 0.502& -0.148&6.454& 24.18 & 22.88& 5.34& 8.18 & 9.97& 14.25\\
$^{286}$114& 2048.6& 2051.6& 2046.0& 0.001& 0.158& -0.347&6.246& 25.21 & 23.32& 1.73& 10.37& 7.57& 26.82\\
$^{287}$114& 2054.9& 2057.6& 2054.8& 0.001& 0.151& -0.173&6.249& 24.84 & 23.22& 2.39& 10.56& 8.18& 17.80\\
$^{288}$114& 2061.1& 2063.1& 2061.3& 0.002& 0.506& -0.184&6.475& 24.10 & 22.91& 4.86& 5.12 & 7.17& 14.84\\
$^{289}$114& 2067.1& 2069.1& 2067.3& 0.002& 0.508& -0.191&6.482& 23.24 & 22.60& 4.66& 7.25 & 6.56& 14.98\\
$^{296}$114& 2107.5& 2107.5& 2105.1& 0.000& 0.538& -0.206&6.257& 10.22 & 18.10& 5.37& 7.44 & 3.08& 15.19\\
$^{298}$114& 2118.0& 2117.9& 2114.5& 0.000& 0.573& -0.316&6.265& 1.31  & 12.94& 5.53& 8.09 & 9.02& 29.45\\
$^{287}$115& 2051.7& 2053.8& 2051.5& 0.001& 0.159& -0.104&6.257& 25.91 & 23.67& 1.01& 9.99 & 17.34& 18.44\\
$^{288}$115& 2058.2& 2060.1& 2057.9& 0.001& 0.152& -0.180&6.259& 25.52 & 23.54& 1.81& 10.25&  7.98& 15.41\\
$^{289}$115& 2064.5& 2066.1& 2064.5& 0.002& 0.143& -0.187&6.261& 24.84 & 23.31& 2.69& 10.65&  7.24& 15.47\\
$^{290}$115& 2070.6& 2072.4& 2070.7& 0.002& 0.515& -0.195&6.499& 23.96 & 22.98& 4.05& 3.20 &  6.64& 15.61\\
$^{291}$115& 2077.3& 2078.5& 2076.8& 0.073& 0.518& -0.213&6.507& 13.28 & 18.94& 3.92& 5.22 &  6.39& 16.10\\
\hline
\end{tabular}
continue...
\end{table*}

\begin{table*}
\renewcommand{\tabcolsep}{0.08cm}
\renewcommand{\arraystretch}{0.5}
\begin{tabular}{|c|c|c|c|c|c|c|c|c|c|c|c|c|c|}
\hline
Nuclei&\multicolumn{3}{c|}{BE}&\multicolumn{3}{c|}{$\beta_2$}&$r_{ch.}$&\multicolumn{2}{c|}{sph.}
&\multicolumn{2}{c|}{prol.}&\multicolumn{2}{c|}{obl.}\\
\hline
&sph.&prol.&obl.&sph.&prol.&obl.&&(DF)$_n$&(DF)$_p$&(DF)$_n$&(DF)$_p$&(DF)$_n$&(DF)$_p$\\
\hline
$^{290}$116& 2067.5& 2068.1& 2067.2& 0.012& 0.144& -0.187&6.271& 25.22 & 23.65& 2.20& 10.26&  7.53& 16.11\\
$^{291}$116& 2073.9& 2075.3& 2073.7& 0.044& 0.522& -0.194&6.561& 20.61 & 22.01& 3.40& 3.65 &  6.92& 16.17\\
$^{292}$116& 2080.3& 2081.7& 2079.9& 0.052& 0.525& -0.212&6.524& 17.45 & 19.62& 3.31& 3.41 &  6.82& 16.53 \\
$^{293}$116& 2086.6& 2087.8& 2086.2& 0.053& 0.529& -0.225&6.532& 16.60 & 20.59& 3.32& 3.29 &  6.81& 16.72\\
$^{293}$117& 2082.8& 2084.4& 2080.2& 0.000& 0.531& -0.355&6.540& 24.50 & 23.45& 2.95& 2.06 &  7.83& 28.12\\
$^{294}$117& 2089.4& 2090.7& 2086.6& 0.043& 0.535& -0.353&6.550& 19.07 & 21.64& 3.00& 2.40 &  7.77& 28.33\\
$^{297}$117& 2107.5& 2108.9& 2105.3& 0.006& 0.550& -0.352&6.580& 19.05 & 21.84& 3.75& 3.33 &  7.62& 28.81\\
$^{294}$118& 2106.2& 2095.8& 2094.2& 0.001& 0.529& -0.235&6.532& 15.06 & 20.13& 4.20& 5.02 &  5.79& 15.99\\
$^{297}$118& 2111.9& 2112.2& 2109.2& 0.000& 0.567& -0.355&6.582& 10.88 & 18.38& 4.78& 5.64 &  4.16& 15.53\\
$^{292}$120& 2061.5& 2060.8& 2057.9& 0.000& 0.547& -0.392&6.281& 28.92 & 26.53& 3.49& 1.47 &  3.51& 21.78\\
$^{293}$120& 2068.7& 2068.3& 2065.1& 0.000& 0.542& -0.388&6.285& 28.56 & 26.49& 3.09& 1.48 &  3.75& 29.11\\
$^{304}$120& 2139.3& 2142.4& 2137.4& 0.000& 0.592& -0.369&6.678& 3.68  & 16.00& 3.44& 0.81 &  8.33& 29.69\\
\hline
\end{tabular}
\end{table*}

\begin{table*}
\caption{The calculated $Q_\alpha$ and life time $T_\alpha$ using SHF(SkI4) and RMF(NL3*) are 
compared with FRDM and experimental data, wherever available.}
\renewcommand{\tabcolsep}{0.08cm}
\renewcommand{\arraystretch}{0.5}
\begin{tabular}{|c|c|c|c|c|c|c|c|c|}
\hline\hline
Nuclei&\multicolumn{4}{c|}{$Q_{\alpha}$ (MeV)}& \multicolumn{4}{c|}{$T_{\alpha}$} \\
\hline
& SHF & RMF & FRDM & Expt. & SHF & RMF & FRDM& Expt. \\
\hline
$^{266}$Db &7.64  &7.17  &7.36  & &1.03$\times10^6$s &8.87$\times10^7$s &10$^{7.15}$s   & \\   
$^{267}$Db &7.41  &7.15  &7.09  & &3.94$\times10^6$s &4.84$\times10^7$s &10$^{7.96}$s   &  \\  
$^{268}$Db &8.01  &7.49  &7.19  & &40673.0s &4.23$\times10^6$s &10$^{7.87}$s   
&   \\             
$^{269}$Db &8.41 &7.98 &7.59 & &717.39s &2.32$\times10^4$s &10$^{6.95}$s  &\\
$^{270}$Db &8.38  &8.19  &7.84  &8.80$\pm$0.1  &1995.7s  &9073.49s &10$^{5.25}$s&1.3min  \\             
$^{258}$Sg &10.02 &9.48  &9.45  &  &4.05ms &0.13s &10$^{-0.83}$s  &  \\ 
$^{260}$Sg &9.49  &9.33  &9.93  &  &0.12s  &0.34s &10$^{-2.15}$s  &  \\         $^{262}$Sg &8.80  &9.04  &9.61  &  &14.42s &2.63s &10$^{-1.26}$s  &  \\ 
$^{271}$Sg &8.84  &8.88  &8.40  &8.54$\pm$0.08 &125.06s&95.39s    &
10$^{3.55}$s   &1.9$^{+2.4}_{-0.6}$s\\  
$^{270}$Bh &8.80  &8.92  &8.18  &8.93$\pm$0.08 &427.56s &173.58s  &10$^{4.76}$s &87.98s \\ 
$^{272}$Bh &9.20  &9.61  &8.88  &9.01 &23.83s &1.49s &10$^{2.36}$s&11.8s \\ 
$^{274}$Bh &8.60  &8.34  &8.71  &8.80$\pm$0.1 &1953.31s &14779.07s &10$^{2.93}$s&1.3min \\ 
$^{264}$Hs &10.20 &10.06 &10.57 & &5.94ms &14.3ms &10$^{-3.18}$s & \\ 
$^{268}$Hs &9.12  &9.59  &9.00  & &7.16s  &28.2ms &10$^{1.24}$s  & \\ 
$^{270}$Hs &8.58  &9.44  &8.69  & &406.49s&79.2ms &10$^{2.23}$s  & \\
$^{272}$Hs &9.98  &10.08 &9.20  & &22.95ms&12.1ms &10$^{0.61}$s  & \\
$^{275}$Hs &9.00  &8.75  &9.39  &9.30$\pm$0.06 &198.15s &1255.21s&10$^{1.09}$s  &0.19$^{+0.22}_{-0.07}$s\\
$^{274}$Mt &10.70 &10.13 &10.06 &10.02$\pm$1.08&8.48ms &249.99ms &10$^{-0.41}$s &472.6ms \\
$^{275}$Mt &10.34 &9.60  &10.06 &9.67 &0.03 s &3.30s &10$^{-0.77}$s &0.98s \\
$^{276}$Mt &10.00 &8.92  &9.93  &9.71 &0.55s  &906.38s &10$^{-0.05}$s&0.72s\\
$^{278}$Mt &9.30  &8.45  &9.37  &9.55$\pm$0.19 &57.68s &34927.55s &10$^{1.54}$s &11.0s \\
$^{270}$Ds &11.38 &10.41 &10.31 & &3.19$\times10^5$s &7.1ms &10$^{-1.89}$s & \\
$^{279}$Ds &10.10 &8.85  &9.68  &9.70$\pm$0.06 &55.67ms &3047.07s &10$^{0.91}$s &0.2$^{+0.05}_{-0.04}$s \\
$^{281}$Ds &13.34 &8.37  &8.55  &8.83 &4.21$\times10^{-8}$s &1.39$\times10^5$s 
&10$^{4.52}$s &1.6min  \\
$^{277}$Rg &12.04 &10.04 &11.50 & &1.31$\times10^{-5}$s &0.865s &10$^{-3.72}$s  & \\
$^{278}$Rg &11.70 &9.89  &11.39 &10.69$\pm$0.08&0.15s&4.93s &10$^{3.11}$s &6.2ms\\
$^{279}$Rg &11.30 &9.52  &10.92 &10.37 &0.56s &27.53s &10$^{-2.36}$s &0.17s \\
$^{280}$Rg &10.80 &9.20  &10.13 &9.75 &19.59s &557.61s&10$^{0.03}$s  &5.1s  \\
$^{281}$Rg &14.77 &12.99 &9.37  &              &1.55$\times10^{-10}$s&1.64$\times10^{-7}$s&10$^{1.90}$s &                   \\
$^{282}$Rg &14.60 &8.76  &8.79  &9.00$\pm$0.1  &6.29$\times10^{-10}$s&15526.06s         &10$^{4.10}$s   &0.74s                 \\
$^{282}$112&15.87 &9.37  &9.42  &              &1.04$\times10^{-12}$s&29.46s            &10$^{1.30}$s   &                      \\
$^{283}$112&14.10 &9.09  &9.01  &9.34$\pm$0.06 &6.32$\times10^{-9}$s &2666.54s          &10$^{3.66}$s   &3.8$^{+1.2}_{-0.7}$s  \\
$^{284}$112&11.48 &10.57 &8.69  &9.17$\pm$0.05 &7.15$\times10^{-5}$s &0.012s            &10$^{3.67}$s   &9.8$^{+18}_{-3.8}$s   \\
$^{285}$112&8.00  &9.98  &8.59  &9.15          &1.97$\times10^7$s    &5.44s             &10$^{5.09}$s   &29s                   \\
$^{294}$112&6.12  &7.71  &7.53  &              &1.69$\times10^{15}$s &2.46$\times10^7$s &10$^{8.15}$s   &                     \\        
$^{282}$113&11.12 &9.89  &9.78  &10.62$\pm$0.08&8.20ms               &21.28s            &10$^{1.66}$s   &88.9ms                \\
$^{284}$113&10.50 &9.68  &9.15  &9.97$\pm$0.05 &46.84ms              &94.43s            &10$^{3.61}$s   &1.36s                 \\
$^{285}$113&5.60  &9.45  &8.97  &9.48$\pm$0.11 &5.74$\times10^{19}$s &203.59s           &10$^{3.86}$s   &7.9s                  \\
$^{286}$113&4.00  &10.34 &8.90  &9.63$\pm$0.1  &9.55$\times10^{33}$s &1.27s             &10$^{4.44}$s   &28.3s                 \\
$^{286}$114&7.60  &10.05 &9.39  &10.19$\pm$0.06&4.57$\times10^8$s    &1.26s             &10$^{2.08}$s   &0.13$^{+0.01}_{-0.01}$s\\
$^{287}$114&8.80  &9.94  &9.31  &9.54          &1.23$\times10^5$s    &29.91s            &10$^{3.39}$s   &3.8s                  \\
$^{288}$114&7.70  &8.61  &9.16  &9.84$\pm$0.05 &1.71$\times10^8$s    &4.83$\times10^4$s &10$^{2.80}$s   &1.9$^{+3.8}_{-0.8}$s  \\
$^{289}$114&7.60  &8.66  &8.87  &9.82          &5.33$\times10^9$s    &3.67$\times10^5$s &10$^{4.86}$s   &2.6s                  \\
$^{296}$114&6.75  &8.32  &8.68  &              &4.68$\times10^{12}$s &5.864$\times10^5$s&10$^{4.43}$s   &                      \\
$^{298}$114&15.6  &7.91  &8.91  &              &6.92$\times10^{-12}$s&2.389$\times10^7$s&10$^{3.63}$s   &                      \\
$^{287}$115&7.09  &11.01 &10.25 &10.59         &1.49$\times10^{12}$s &42.8ms            &10$^{0.63}$s   &32ms                  \\
$^{288}$115&10.80 &10.75 &10.12 &10.48         &318.88ms             &426.6ms           &10$^{1.33}$s   &250ms                 \\
$^{289}$115&10.60 &10.74 &10.03 &10.31$\pm$0.09&482.31ms             &204.84ms          &10$^{1.26}$s   &320ms                 \\
$^{290}$115&12.20 &9.23  &9.92  &9.95$\pm$0.4  &0.165s               &11050.16s         &10$^{1.90}$s   &0.023s                \\
$^{291}$115&8.86  &10.15 &9.66  &              &87783.45s            &8.45s             &10$^{2.34}$s   &                      \\
\hline
\end{tabular}
continue...
\end{table*}

\begin{table*}
\renewcommand{\tabcolsep}{0.08cm}
\renewcommand{\arraystretch}{0.5}
\begin{tabular}{|c|c|c|c|c|c|c|c|c|}
\hline
Nuclei&\multicolumn{4}{c|}{$Q_{\alpha}$ (MeV)}& \multicolumn{4}{c|}{$T_{\alpha}$} \\
\hline
& SHF & RMF & FRDM & Expt. & SHF & RMF & FRDM& Expt. \\
\hline
$^{290}$116&8.90  &11.76 &11.12 &10.84$\pm$0.08&24443.12s            &0.23ms            &10$^{-2.12}$s  &7.1$^{+3.2}_{-1.7}$s  \\
$^{291}$116&8.50  &10.58 &11.11 &10.74$\pm$0.07&7.77$\times10^6$s    &2.25s             &10$^{-1.03}$s  &18$^{+22}_{-6}$ms     \\
$^{292}$116&9.40  &9.689 &10.82 &10.71$\pm$0.15&528.65s              &66.091s           &10$^{-1.37}$s  &33$^{+155}_{-15}$ms   \\
$^{293}$116&9.30  &9.679 &10.94 &10.67$\pm$0.06&12923.99s            &825.50s           &10$^{-0.59}$s  &53$^{+62}_{-19}$ms    \\
$^{293}$117&10.70 &9.989 &11.68 &11.03$\pm$0.08&1.07s                &104.71s           &10$^{-2.40}$s  &21ms                  \\
$^{294}$117&9.40  &10.00 &11.67 &10.81$\pm$0.1 &15027.71s            &213.62s           &10$^{-2.05}$s  &112ms                 \\
$^{297}$117&10.74 &9.52  &11.74 &              &0.844s               &2869.78s          &10$^{-2.55}$s  &                      \\
$^{294}$118&11.10 &9.60  &12.28 &10.84         &33.36ms              &553.44s           &10$^{-4.24}$s  &7.1ms                 \\
$^{297}$118&4.40  &9.95  &12.10 &              &5.13$\times10^{32}$s &555.59s           &10$^{-2.79}$s  &                      \\
$^{292}$120&12.42 &10.37 &13.89 &              &9.68$\times10^{-5}$s &12.30s            &10$^{-6.96}$s  &                      \\
$^{293}$120&12.24 &11.05 &13.69 &              &2.81ms               &2.09s             &10$^{-5.51}$s  &                      \\
$^{304}$120&12.07 &10.19 &13.82 &              &0.58ms               &42.53s            &10$^{-6.83}$s  &                      \\
\hline
\end{tabular}
\end{table*}

\subsection{Shape and shape co-existence}

Nuclei have different binding energy corresponding to different shape 
configuration in the ground or excited states.
In some cases, it so happens that the binding energy of two different
configurations coincide or very close to each other
known as shape co-existence~\cite{rana,raj,sara,egido}.
This phenomenon is more common in superheavy region, leads to 
complex structure of these nuclei which 
provides the information about the oscillation of nuclei 
between two or three existing shape by perturbing small energy.
Here, in our studied nuclei, we find many such examples, where
the ground and first excited binding energies are degenerated.
Also, in several cases the nuclei have well defined three 
distinct solutions (oblate, spherical, prolate) with almost same energy 
which strongly indicates the nuclei may oscillate 
from oblate to spherical to prolate configuration.

For analyzing the result , we took a small binding energy difference around $\leq$ 2 MeV 
to take care of shape co-existence for our calculations.
Because of this small difference in binding energies between these two
solutions, the ground state can be changed to the low-lying excited state or 
vice-versa by a small change in input parameters like pairing gaps etc.
It is seen from Tables I and II the shape co-existence is observed 
for  $^{277, 278, 281, 282}$Rg, $^{282, 283, 284, 285}$112, 
$^{286, 287, 288}$114 $^{294, 297}$118 and $^{292, 293}$120 in SHF calculations   
and RMF also produces the shape co-existence for $^{294}$112, $^{286}$113, 
$^{288, 289, 296, 298}$114, $^{290, 291}$115, $^{291, 292}$116, 
$^{293, 294, 297}$117, $^{297}$118, $^{292, 293}120$.
A shape co-existence of nuclei indicates that there is a competition 
among the different shapes to acquire the possible ground 
state for maximum stability and the final shape could be an admixture
of these low-lying bands.

\subsection{Density and bubble structure}

\begin{figure}[h]
\vspace{0.75cm}
\includegraphics[width=0.95\columnwidth]{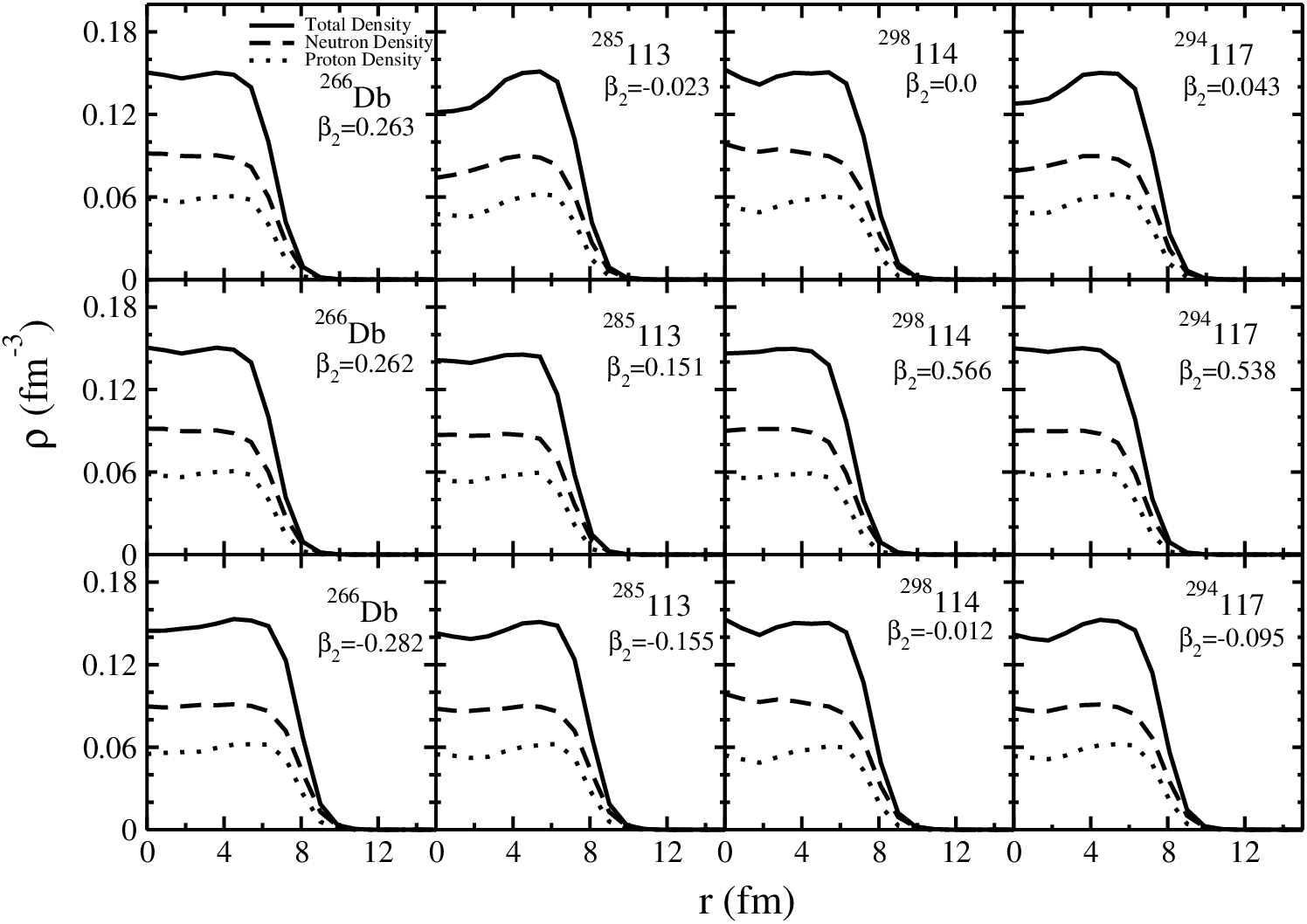}
\caption{\label{dshf}The neutron, proton and total matter density 
distribution for some of the selected superheavy nuclei using
SHF(SkI4).}
\end{figure}

\begin{figure}[h]
\vspace{0.75cm}
\includegraphics[width=0.95\columnwidth]{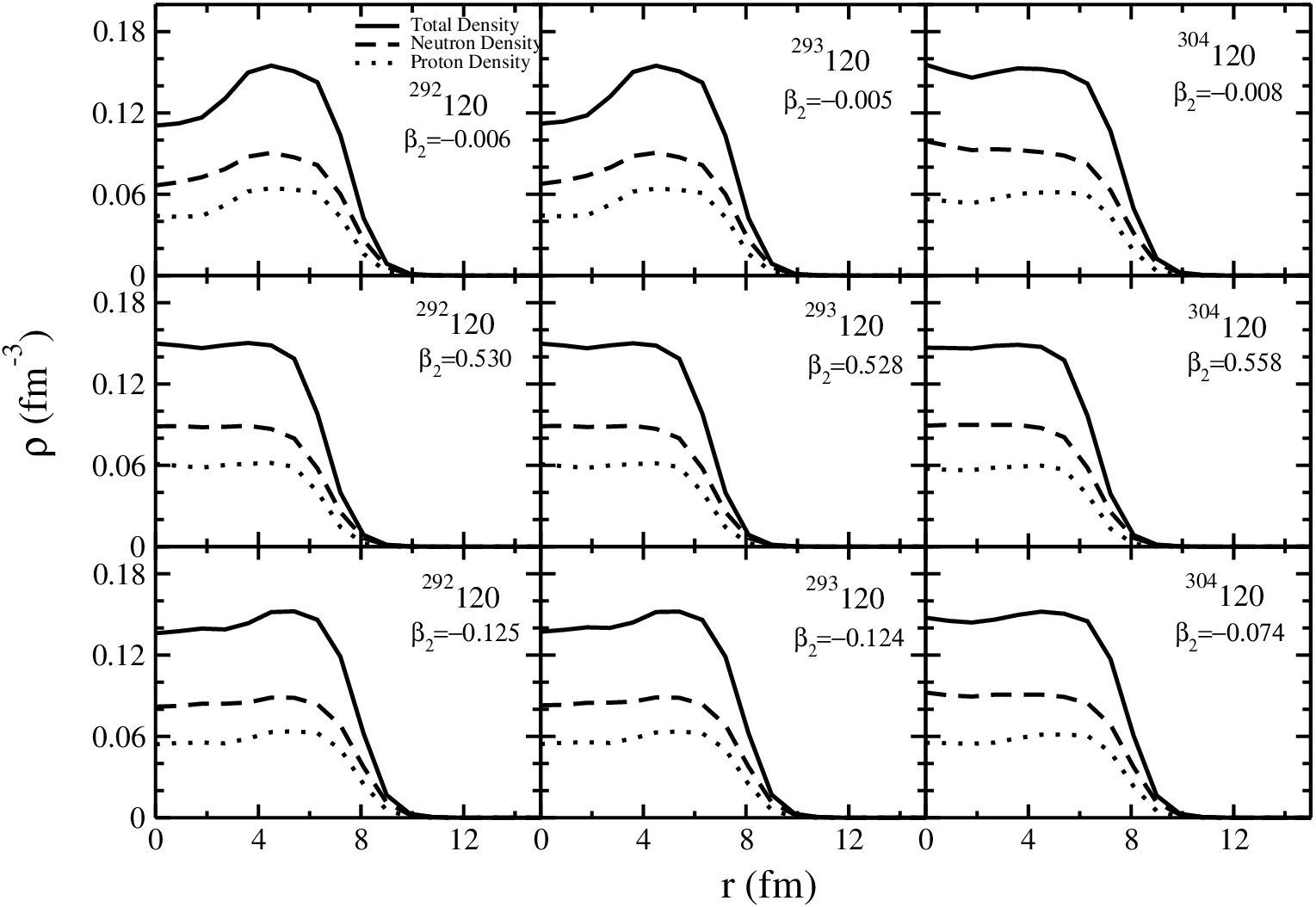}
\caption{\label{dshf120}Same as Fig. 3, but for $^{292,293,304}$120 with SHF(SkI4).}
\end{figure}

\begin{figure}[h]
\vspace{0.75cm}
\includegraphics[width=0.95\columnwidth]{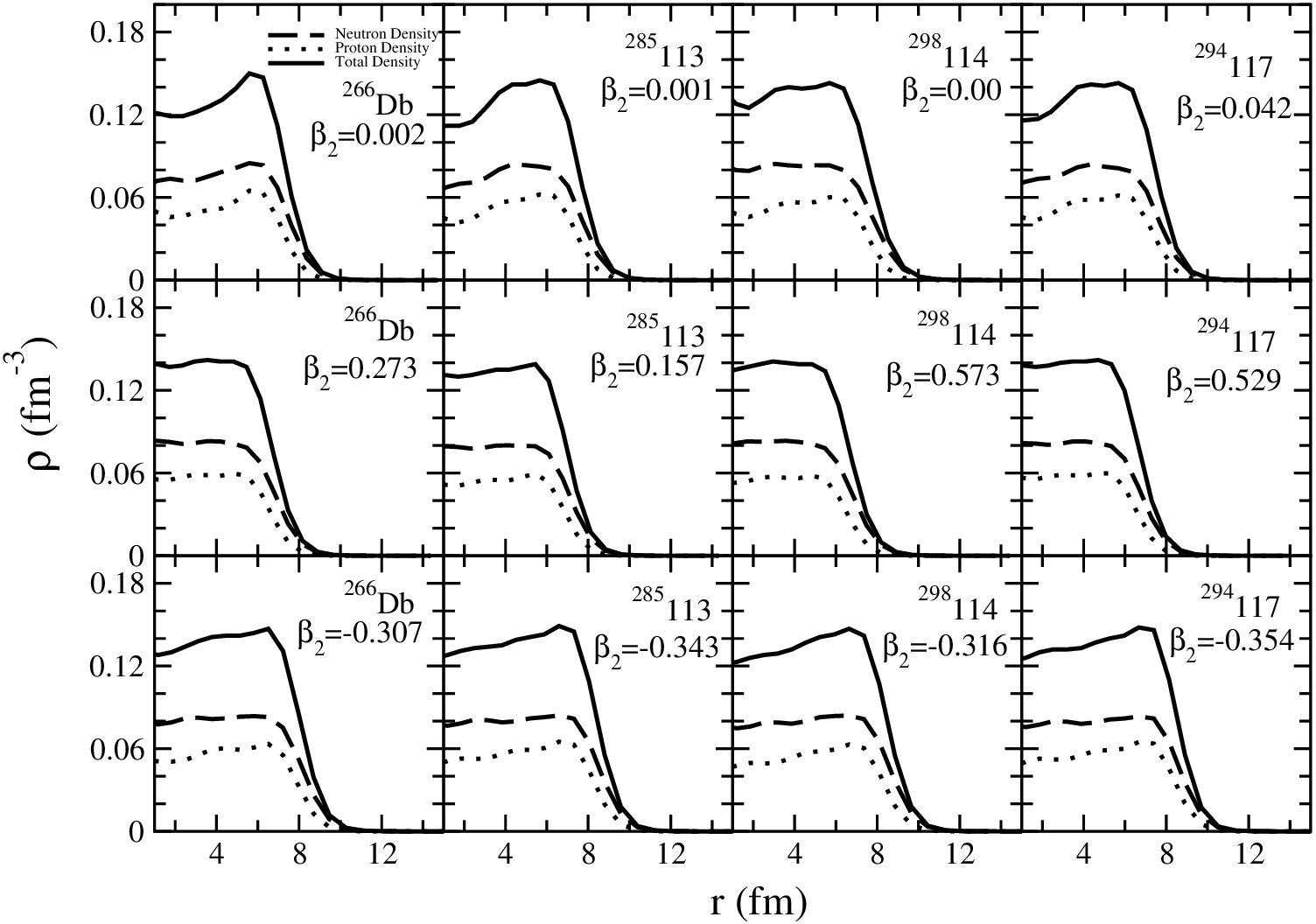}
\caption{\label{dnl3} Same as Fig. 3, but with RMF(NL3*).}
\end{figure}

\begin{figure}[h]
\vspace{0.75cm}
\includegraphics[width=0.95\columnwidth]{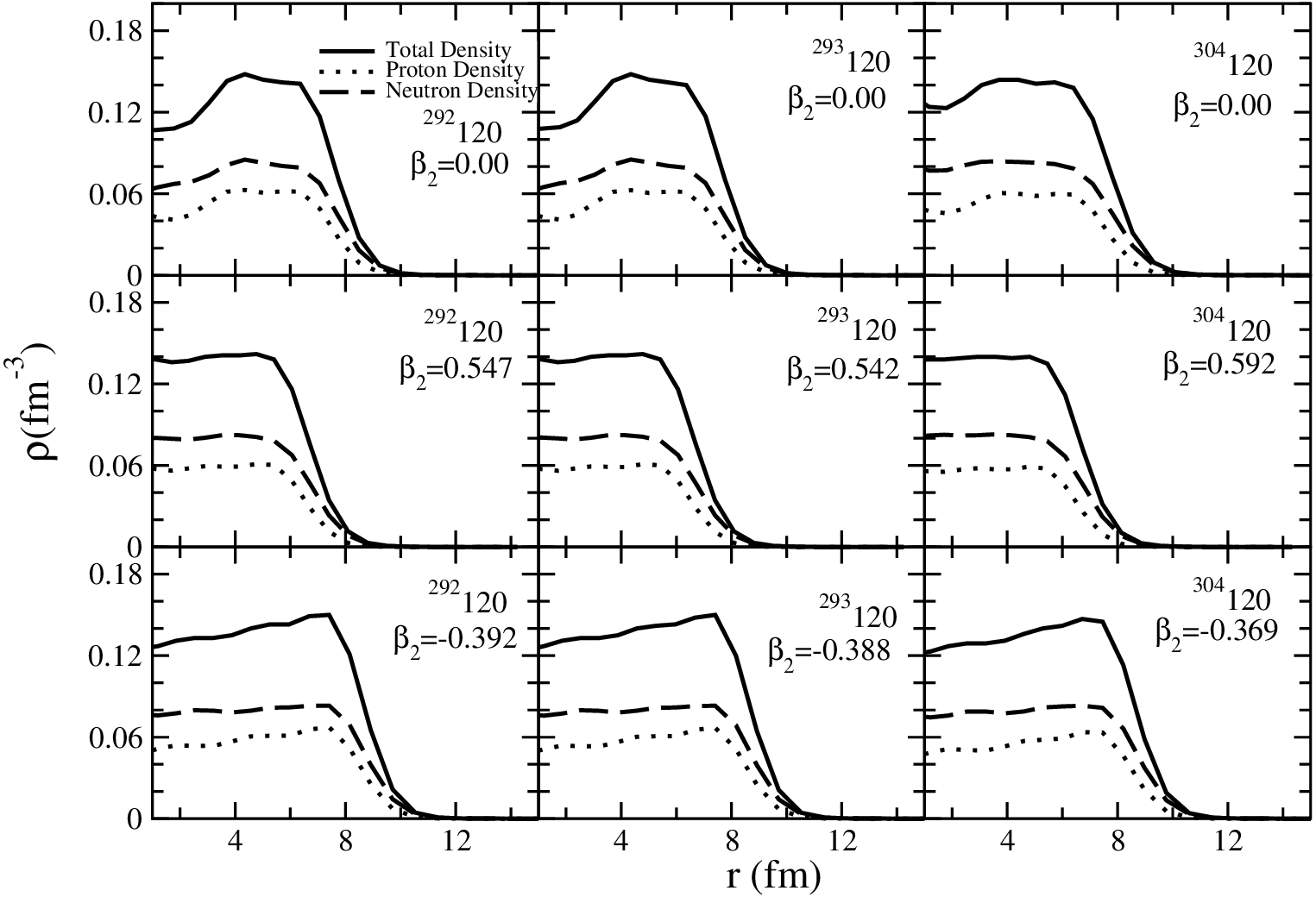}
\caption{\label{dnl3120}Same as Fig. 3, but for $^{292,293,304}$120 with RMF(NL3*).}
\end{figure}

\begin{figure}[h]
\vspace{0.75cm}
\includegraphics[width=0.95\columnwidth]{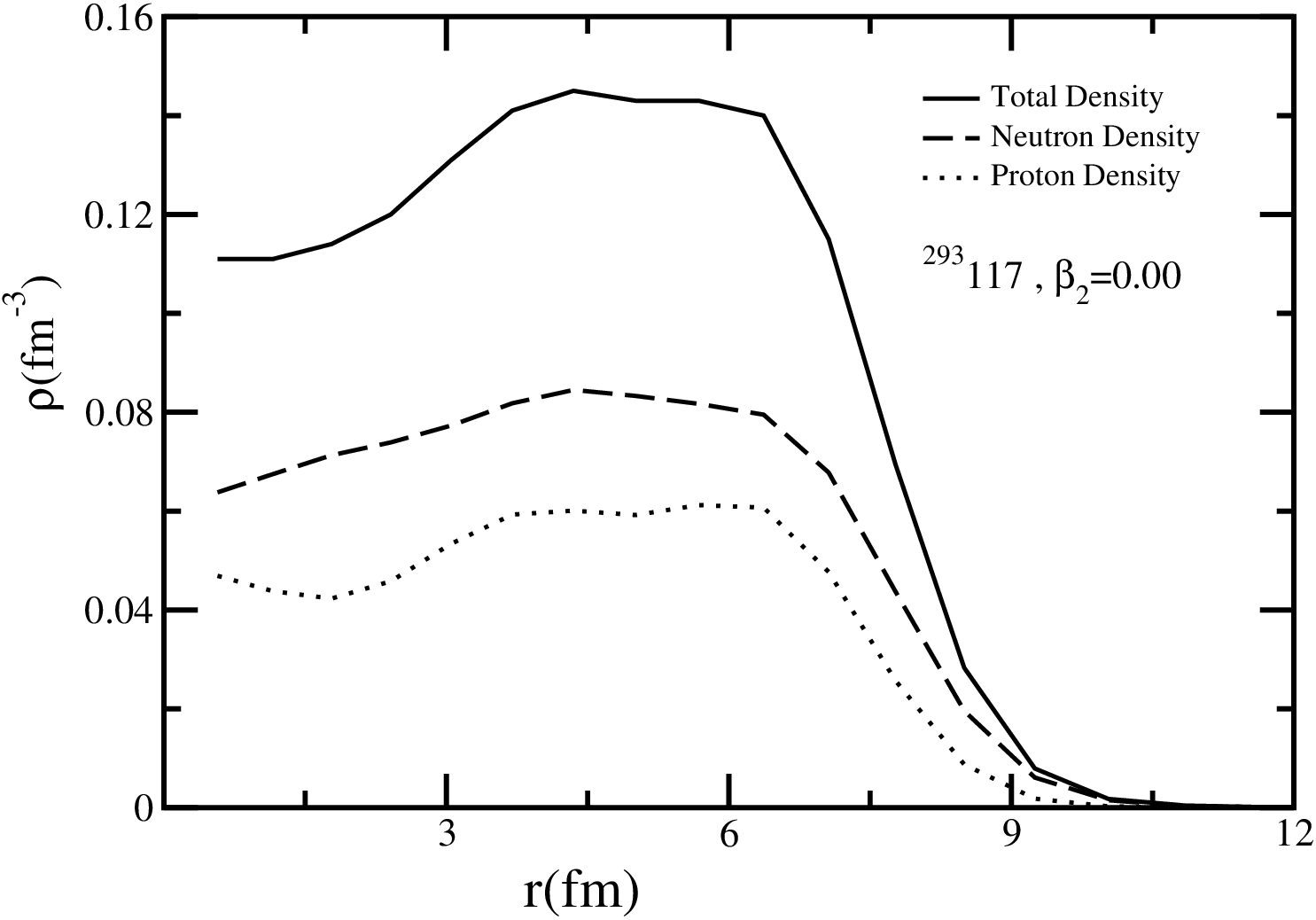}
\caption{\label{dnl3117}The neutron, proton and total matter density 
distribution for $^{293}$117 with RMF(NL3*).}
\end{figure}

The density of a nucleus has the gross information about the size, 
shape and distribution of nucleons. 
The density distribution from SHF(SkI4) and RMF(NL3*) for 
some of the selected nuclei are given in Figures ~\ref{dshf} to ~\ref{dnl3117} obtained
for all the three solutions, i.e. spherical, prolate and oblate.
Normally, the density at the center of nucleus has the maximum 
value and decays to zero at the surface. 
However, in some specific cases, e.g. for spherical solutions
of $^{285}$113, $^{294}$117, $^{292,293}$120, this trend of density distribution
shows an anomalous behavior, i.e. a dip at the center and a hump
nearby to it following a slow decreasing in density to zero at the
surface. This type of density distribution is known as
bubble structure ~\cite{wheeler}.
The first possibility for existence of toroidal or bubble structure 
other than the spherical topological structure was 
suggested by Wheeler ~\cite{wheeler}.
The occurrence of bubble nuclei has been extensively studied by 
Wilson~\cite{wilson} and later by Siemens and Bethe~\cite{bethe}.  
Several models like independent particle model~\cite{wong} and Hartree 
Fock Model~\cite{davies} also investigated the possibility of low density 
region at the center of the nuclei.
This structure is not confined to a particular region, but have the 
possibility for light to superheavy nuclei~\cite{charge,grasso}.
The contribution of density at {\it r = 0} is offered by non zero wavefunction
of {\it s}-states and depopulation of this level leads to the depletion of
central density and formed a bubble like structure \cite{grasso}.
The inversion of {\it s}-states to other higher states 
may also be the possibility of depletion of central density
and can be interpreted as {\it s-d} orbital inversion 
\cite{zao,khan}.

\newpage
The bubble effect can be realized quantitatively by calculating 
depletion fraction for the neutrons as well as protons using the
relation ~\cite{grasso} 
\begin{equation}
(D.F.)_{\alpha} = \frac{(\rho_{max})_{\alpha} - (\rho_{cen})_{\alpha}}{(\rho_{max})_{\alpha}}\times100,
\end{equation}
where $\rho_{max}$, $\rho_{cen}$ represent the  maximum 
and central density respectively and $\alpha$ denotes neutron or proton.
The calculated values of depletion fraction for neutron as 
well as proton by using non-relativistic and relativistic 
densities are given in Tables I and II for all three solutions.
In our present study, most of the nuclei achieve the prolate shape 
as their ground state solution both in the SHF and RMF calculations.
Here, we do not find significant bubble structure, i.e. 
no good amount of depletion fraction are estimated for this shape, except a few cases.
It is evident from Tables I and II, a remarkable depletion fraction is 
there in $^{292,293}$120 over the whole series of nuclei 
on spherical solution in both models.

\subsection{$Q_{\alpha}$ and $T_{\alpha}$ for superheavy nuclei}

The $Q_{\alpha}$ value of a nucleus provides valuable information
for its stability. This is more important for superheavy nuclei,
which is the central theme of study in the present paper. Thus, it
is worthwhile to evaluate the $Q_{\alpha}$ values for the nuclei
which will also give us to estimate the life time $T_\alpha$ of the considered
nuclei. It is to be noted that mostly, $\alpha-$decay is the decay mode 
for these synthesized  elements, which end with spontaneous fission.
Recently, it is predicted that the $\beta-$decay may play an important
role for some of the SHEs \cite{karpo12}.
The calculated $Q_\alpha$ and  $T_\alpha$  for 
$Z$ = 105 - 120 within the relativistic and non-relativistic formalism 
using NL3* and SkI4 parameters are listed in Table III. 
The $Q_\alpha$ is calculated by using the following expression
\begin{equation}
Q_{\alpha} = BE(N,Z) - BE(N-2,Z-2) - BE(2,2).
\end{equation}
Here, BE(N,Z) and BE(N-2,Z-2) are the binding energy of the parent 
and daughter nuclei, respectively. $BE(2,2)$ is the binding energy of 
$\alpha$ particle ($^4He$) i.e. $28.3$ MeV~\cite{audi03}. 
We are using the empirical Viola-Seaborg formula~\cite{viola} for 
the calculation of $T_\alpha$ which is define as:
\begin{equation}
\log_{10}(T_{\alpha})=\frac{(aZ + b)}{\sqrt Q_\alpha}  + (cZ + d )  + h\log,
\end{equation}
where Z is the charge number of the parent nucleus and $a$, $b$, $c$, $d$
are the fitting constants defined as~\cite{sobi89} 
$a$ = 1.66175, $b$ = -8.5166, $c$ = -0.20228, $d$ = -33.9069.
The last term in above equation is known as even-odd hindrance factor 
and is obtained by fits to odd nuclei. 
It takes the value as $hlog = 0.0 (for Z = even, N = even),\; 
hlog = 0.772 (for Z = odd, N = even),\; hlog = 1.066(for Z = even, N = odd),\; 
hlog = 1.114 (for Z = odd, N=odd)$.
From the above equation, one can find that $T_{\alpha}\propto10^{\frac{1}{\sqrt{Q}}}$, 
hence a small change in $Q_{\alpha}$ creates a large difference 
in $T_{\alpha}$. This observation reflects in our results (see Table III).

\section{Summary}

In summary, we study a broad spectrum of the superheavy nuclei for $Z$ = 105 - 120 
in the frame work of relativistic and nonrelativistic mean field formalisms. The
calculations are done in an axially deformed coordinates using constant gap 
BCS pairing. Except $Z$=120, all other isotopes are synthesized artificially 
whose properties are studied.
From the calculated binding energies, $Q_{\alpha}$ and $T_{\alpha}$ are estimated. 
The results are compared with the most acceptable FRDM data as well as with 
the experimental observations, wherever available.
Over all comparison of the results are quite agreeable with each other in all the three models
along with the experimental data.

On the basis of the depletion of central density, a bubble structure is  
studied for some superheavy nuclei in the excited states. This observation
is model independent. The maximum depletion fraction calculated in the 
considered nuclei is for $^{292,293}$120.
In some of the cases, [Fig.~\ref{dnl3117}] a depletion of proton density nearer 
to the center (but not at the center exactly) indicates a special type of proton 
distribution. In this case, the center is slightly bulgy
and a considerably depression afterward. Again a big hump almost at the
mid distance of the center and the surface appears in the density distribution
for the spherical state. This feature of the nucleus is unique and need more
investigation. Work in this direction is in progress \cite{singh12}.

\section{Acknowledgments}
One of the author (MI) acknowledge the hospitality provided by Institute 
of Physics, Bhubaneswar during the work.

\end{document}